\documentclass[apjl]{emulateapj}
\begin{document}
\title{The primordial abundance of $^4$He: evidence for non-standard
big bang nucleosynthesis}

\author{Yuri I. Izotov}

\affil{Main Astronomical Observatory, Ukrainian National Academy of Sciences,
27 Zabolotnoho str., Kyiv 03680, Ukraine}
\email{izotov@mao.kiev.ua}
\author{Trinh X. Thuan}

\affil{Astronomy Department, University of Virginia, P.O. Box 400325, 
Charlottesville, VA 22904-4325}
\email{txt@virginia.edu}

\begin{abstract}
We present a new determination of the primordial
helium mass fraction $Y_p$, based on 93 
spectra of 86 low-metallicity extragalactic H {\sc ii} regions, 
and taking into account the latest developments concerning systematic effects. 
These include collisional and fluorescent
enhancements of He {\sc i} recombination lines, underlying He {\sc i} 
stellar absorption lines, collisional and fluorescent 
excitation of hydrogen lines and 
temperature and ionization structure of the H {\sc ii} region. 
Using Monte Carlo methods to solve simultaneously for the above 
systematic effects, we find the best value to be  
$Y_p$ = 0.2565 $\pm$ 0.0010(stat.) $\pm$ 0.0050(syst.).
This value is higher at the 2$\sigma$ level than the value given 
by Standard Big Bang Nucleosynthesis (SBBN), implying 
deviations from it. The effective number of light neutrino species $N_\nu$ 
is equal to 3.68$^{+0.80}_{-0.70}$ (2$\sigma$) and 
3.80$^{+0.80}_{-0.70}$ (2$\sigma$) for a neutron lifetime
$\tau_{\rm n}$  equal to 885.4 $\pm$ 0.9 s and 878.5 $\pm$ 0.8 s, respectively,
i.e. it is larger than the experimental value of 2.993$\pm$0.011.  
\end{abstract}


\section{INTRODUCTION}\label{intro}
The determination of the primordial $^4$He (hereafter He) abundance and of 
some 
other light elements such as D, $^3$He and $^7$Li, 
plays an important role in testing
cosmological models.
In the standard theory of big bang nucleosynthesis (SBBN), given the 
number of light neutrino species, the abundances of these light elements 
depend only on one cosmological parameter, the   
baryon-to-photon number ratio $\eta$.

Because of the strong dependence of its abundance on $\eta$, deuterium has
become the baryometer of choice. The D/H measurements in damped Ly$\alpha$ 
systems appear to converge to a mean primordial value 
log D/H = $-$4.56 $\pm$ 0.04, corresponding to a  baryon mass fraction 
$\Omega_bh^2$ = 0.0213 $\pm$ 0.001 \citep{P08}.
This estimate of $\Omega_bh^2$
is in excellent agreement with the value of 0.02273 $\pm$ 0.00062 obtained 
by \citet{D09} from analysis of five years of observations with the
Wilkinson Microwave Anisotropy Probe (WMAP).

Although He is not a sensitive baryometer ($Y_p$ depends only 
logarithmically on the baryon density), its primordial abundance depends much 
more sensitively than that of D on the expansion rate of the Universe and 
on a possible lepton asymmetry in the early universe 
\citep{St06,St07}. Thus, accurate measurements of the 
primordial abundance of He are required to check the consistency of SBBN. 

However, to detect small deviations from SBBN and 
make cosmological inferences, $Y_p$ has to be determined 
to a level of accuracy of less than a few percent.
The primordial abundance of He can, in principle, be derived 
accurately from observations of the He and 
H emission lines from low-metallicity H {\sc ii} regions.
Several groups have used this technique to derive the primordial  
He mass fraction $Y_p$, with somewhat different results.
In the most recent study \citet{I07} based on a large sample
of 93 spectra of 86 low-metallicity extragalactic H {\sc ii} regions
(hereafter the HeBCD sample) derived $Y_p$ = 0.2472 $\pm$ 0.0012 and 
$Y_p$ = 0.2516 $\pm$ 0.0011, using \citet{B99,B02} and \citet{P05} He {\sc i}
emissivities, respectively. On the other hand, \citet{P07} obtained 
$Y_p$ = 0.2477 $\pm$ 0.0029 based on a sample of 5 H {\sc ii} regions,
\citet{P05} He {\sc i} emissivities, and adopting non-zero temperature
fluctuations. \citet{F06} derived $Y_p$ = 0.250 $\pm$ 0.004 for a sample
of 31 H {\sc ii} regions studied 
by \citet{IT04}, adopting \citet{B99,B02} He {\sc i}
emissivities.

It is generally believed that the accuracy of the determination of the 
primordial He abundance is limited presently, not so much 
by statistical uncertainties, but by our ability to account for systematic 
errors and biases.   
There are many known effects we need to correct for  
to transform the observed He {\sc i} line intensities into a He abundance. 
These effects are: (1) reddening, (2) underlying stellar
absorption in the He {\sc i} lines, (3) collisional 
excitation of the He {\sc i} lines which make their intensities 
 deviate from their recombination values, 
(4) fluorescence of the He {\sc i} lines which also make their intensities 
deviate from their 
recombination values, (5) collisional and (6) fluorescent excitation of 
the H lines, 
(7) the temperature 
structure of the H {\sc ii} region and (8) its ionization structure. All
these corrections are at a level of a few percent 
except for effect (3) that can be much higher, exceeding 10\% in the case
of the He {\sc i} $\lambda$5876 emission line in hot 
and dense H {\sc ii} regions. Most of these effects were analyzed and 
taken into account by \citet{I07} for their HeBCD sample.
 
We present here a new determination of the primordial He abundance. 
A new study is warranted because there are several  
recent new developments that allow more accurate estimates of some of the 
systematic effects mentioned above.  
Thus, \citet{G05} have calculated evolutionary
stellar population synthesis models with high 
spectral resolution, so that equivalent
widths of H {\sc i} and He {\sc i} absorption lines for different ages of
single stellar populations and for a wide range of metallicities 
are now available.
Furthermore, \citet{L09a} has 
estimated the collisional enhancement of hydrogen
Balmer lines, while \citet{L09b} have 
considered the fluorescent excitation of Balmer
lines in gaseous nebulae (their case D). We incorporate all these 
new calculations in our new determination of the primordial He abundance.

In \S\ref{method}, we briefly discuss  
the method used to derive He abundances in 
individual objects and the primordial He abundance from the total sample.
In \S\ref{primo}, we derive
the best value for $Y_p$ and the linear regression slope d$Y$/d$Z$.
The cosmological implications of our new results are discussed in 
\S\ref{cosmo}. We summarize our conclusions in \S\ref{summary}.

\section{THE METHOD}\label{method}

\subsection{Linear regressions}\label{method1}

As in our previous work \citep[see ][ and references therein]{I07},
we determine the primordial He mass fraction
$Y_p$ by fitting the data points in the $Y$ -- O/H
plane with a linear regression line of the
form \citep{PTP74,PTP76,P92}
\begin{equation}
Y = Y_p + \frac{{\rm d}Y}{{\rm d}({\rm O/H})} ({\rm O/H}),               \label{eq:YvsO}
\end{equation}
where
\begin{equation}
Y=\frac{4y(1-Z)}{1+4y} \label{eq:Y}
\end{equation}
is the He mass fraction, $Z$ is the heavy element mass fraction,
$y$ = ($y^+$ + $y^{2+}$)$\times$$ICF$(He$^+$+He$^{2+}$) is the He 
abundance,
$y^+$ $\equiv$ He$^+$/H$^+$ and $y^{2+}$ $\equiv$ He$^{2+}$/H$^+$ are 
respectively the abundances of singly and doubly ionized He, and 
$ICF$(He$^+$+He$^{2+}$) is the ionization correction factor for He.

The slope of the $Y$ -- O/H linear regression can be written as:
\begin{equation}
\frac{{\rm d}Y}{{\rm d}({\rm O/H})} = 12\frac{{\rm d}Y}{{\rm d}{\rm O}} =
18.2\frac{{\rm d}Y}{{\rm d}Z}, \label{eq:dO}
\end{equation}
where O and $Z$ are respectively the mass fractions of oxygen
and heavy elements.

We also take into account the depletion of oxygen on dust grains. \citet{I06}
demonstrated that the Ne/O abundance ratio for low-metallicity BCDs is
not constant, but increases with increasing oxygen abundance. This
effect is small, with $\Delta$logNe/O = 0.1 when the oxygen abundance
changes from 12+logO/H = 7.0 to 8.6.
We attribute such a change to oxygen
depletion, and correct the oxygen abundance for it, using the 
log Ne/O versus oxygen abundance regression line found by \citet{I06}, and
assuming that depletion is absent in galaxies with 12+logO/H = 7.0.

To derive the parameters of the  $Y$ versus O/H linear regressions,
 we use the maximum-likelihood method \citep{Pr92}
 which takes into account the errors in both $Y$ and O/H for each object.

\subsection{A Monte Carlo algorithm for determining 
the best value of $y^+$}\label{method3}

Following \citet{I07}, we use the five 
strongest He {\sc i} $\lambda$3889, $\lambda$4471, 
$\lambda$5876, $\lambda$6678 and $\lambda$7065 emission lines to derive 
the electron
number density $N_e$(He$^+$) and the optical depth $\tau$($\lambda$3889). 
The He {\sc i} $\lambda$3889 and 
$\lambda$7065 lines play an important role because they are particularly 
sensitive to both quantities. Since the
He {\sc i} $\lambda$3889 line is blended with the H8 $\lambda$3889 line, 
we have subtracted the latter, assuming its intensity to be equal to 0.107 
$I$(H$\beta$) \citep{A84}. 

The derived $y^+$ abundances depend also on a number of  
other parameters: the fraction $\Delta$$I$(H$\alpha$)/$I$(H$\alpha$) of the
H$\alpha$ emission line flux due to collisional excitation, the electron
number density $N_e$(He$^+$), the electron temperature $T_e$(He$^+$), 
the equivalent
widths EW$_{abs}$($\lambda$3889), EW$_{abs}$($\lambda$4471), 
EW$_{abs}$($\lambda$5876), 
EW$_{abs}$($\lambda$6678) and EW$_{abs}$($\lambda$7065) of He {\sc i} stellar 
absorption lines, and the optical depth 
$\tau$($\lambda$3889) of the He {\sc i} $\lambda$3889 emission line. 
To determine the best weighted mean value of $y^+$, 
we use the Monte Carlo procedure described in \citet{I07}, 
randomly varying each of the above parameters within a specified range. 

Additionally, in those cases when the nebular He {\sc ii} $\lambda$4686 
emission line was detected, we have added to $y^+$ 
the abundance of doubly ionized 
helium $y^{2+}$ $\equiv$ He$^{2+}$/H$^+$ \citep{I07}. 

\subsection{Parameter set for the He abundance determination}\label{method4}

For the determination of He abundance, we adopt He {\sc i}
emissivities from \citet{P05} and take into account 
the following systematic effects: 
1) reddening; 2) the temperature structure of the H {\sc ii} region,
i.e. the temperature difference between $T_e$(He$^+$)
and $T_e$(O {\sc iii}); 4) underlying stellar He {\sc i} absorption; 
4) collisional and fluorescent excitation of He {\sc i} lines;
5) collisional and fluorescent excitation of hydrogen lines; and 
6) the ionization structure 
of the H {\sc ii} region. Most of these effects were analyzed by \citet{I07}
(see also references therein).

We define the following set of parameters: 

1. The reddening law of \citet{W58} is adopted. \citet{I07} 
have shown that He abundances are not sensitive to the particular 
reddening law adopted.
For example, use of the \citet{C89} reddening curve 
results in He and other element
abundances similar to those obtained with the \citet{W58} 
reddening law. 
The extinction coefficient $C$(H$\beta$) is derived from the observed hydrogen
Balmer decrement, after correcting the 
H$\alpha$, H$\beta$, H$\gamma$ and H$\delta$ line fluxes 
for the effects of collisional and fluorescent excitation.
Finally, all emission lines are corrected for reddening, adopting the derived
$C$(H$\beta$).

2. The electron temperature
of the He$^+$ zone is varied in the range $T_e$(He$^+$) =
(0.95 -- 1.0)$\times$$T_e$(O {\sc iii}). We have chosen this range 
following the work of 
\citet{G06} and \citet{G07} who have derived the electron temperature
in the H$^+$ zone from the Balmer and Paschen discontinuities in the 
spectra of more than 100 H {\sc ii} regions, 
and showed that $T_e$(H$^+$) differs from 
$T_e$(O {\sc iii}) by not more than 5\%. 
We also assume that 
$T_e$(He$^+$) = $T_e$(H$^+$) because the H$^+$ and He$^+$ zones in our
objects are nearly coincident.

3. Oxygen abundances 
are calculated by considering two possible values of the 
electron temperature: 1) $T_e$ = $T_e$(He$^+$)
and 2) $T_e$ = $T_e$(O {\sc iii}).
 
4. $N_e$(He$^+$) and $\tau$($\lambda$3889) are varied  
respectively in the ranges 10 -- 450 cm$^{-3}$ and 0 -- 5, typical for
extragalactic H {\sc ii} regions.

5. The fraction of H$\alpha$ emission due to collisional excitation is
varied in the range 0\% -- 5\%, in accordance with \citet{SI01} and
\citet{L09a}. The fraction of H$\beta$, H$\gamma$ and H$\delta$ 
emission due to  
collisional excitation is adopted to be 60\% that of the H$\alpha$ emission,
in accordance with \citet{L09a}. We note that \citet{I07} underestimated 
that fraction for H$\beta$, adopting a value of only 1/3, and 
neglected altogether to correct the H$\gamma$
and H$\delta$ emission lines for collisional excitation.   

6. \citet{L09b} have shown that the fraction of H$\beta$ emission due to
fluorescent excitation by the 
far-UV non-ionizing stellar continuum could be as 
high as 2\%,  
and somewhat lower for the H$\alpha$ emission (their case D). 
We have adopted the conservative value of 1\%  for 
the fraction of H$\alpha$, H$\beta$, H$\gamma$ and H$\delta$
emission due to fluorescent excitation, since a similar effect could
affect the He {\sc i} emission lines 
and partly compensate the effect for the Balmer H lines 
(the He abundance is calculated relative to that of H).  

7. The equivalent width of the 
He {\sc i} $\lambda$4471 absorption line is chosen to be 
EW$_{abs}$($\lambda$4471) = 0.4\AA, following \citet{I07} and
\citet{G05}. The equivalent 
widths of the other absorption lines are fixed according to the ratios  
EW$_{abs}$($\lambda$3889) / EW$_{abs}$($\lambda$4471) = 1.0,
EW$_{abs}$($\lambda$5876) / EW$_{abs}$($\lambda$4471) = 0.8,
EW$_{abs}$($\lambda$6678) / EW$_{abs}$($\lambda$4471) = 0.4 and 
EW$_{abs}$($\lambda$7065) / EW$_{abs}$($\lambda$4471) = 0.4. 
The EW$_{abs}$($\lambda$5876) / EW$_{abs}$($\lambda$4471) and 
EW$_{abs}$($\lambda$6678) / EW$_{abs}$($\lambda$4471) ratios 
were set equal to the values predicted for these ratios by   
a Starburst99 \citep{L99} instantaneous
burst model with an age 3-4 Myr and a heavy element mass fraction
$Z$ = 0.001 -- 0.004, 0.8 and 0.4 respectively. 
These values are significantly higher 
than the corresponding ratios of 0.3 and 0.1 adopted by \citet{I07}. 
We note that the value chosen for the 
EW$_{abs}$($\lambda$5876) / EW$_{abs}$($\lambda$4471) ratio is also 
consistent with the one given by \citet{G05}. 
Since the output high-resolution spectra in Starburst99 are calculated only
for wavelengths $\leq$ 7000\AA, we do not have a prediction for the 
EW$_{abs}$($\lambda$7065) / EW$_{abs}$($\lambda$4471) ratio. We set it to 
be equal to 0.4, the value of 
the EW$_{abs}$($\lambda$6678) / EW$_{abs}$($\lambda$4471) ratio. 

8. The He ionization correction factor $ICF$(He$^+$+He$^{++}$) is adopted from
\citet{I07}. 

\section{The primordial He mass fraction $Y_p$ and the slope d$Y$/d$Z$}\label{primo}

\begin{figure*}
\figurenum{1}
\epsscale{1.1}
\plottwo{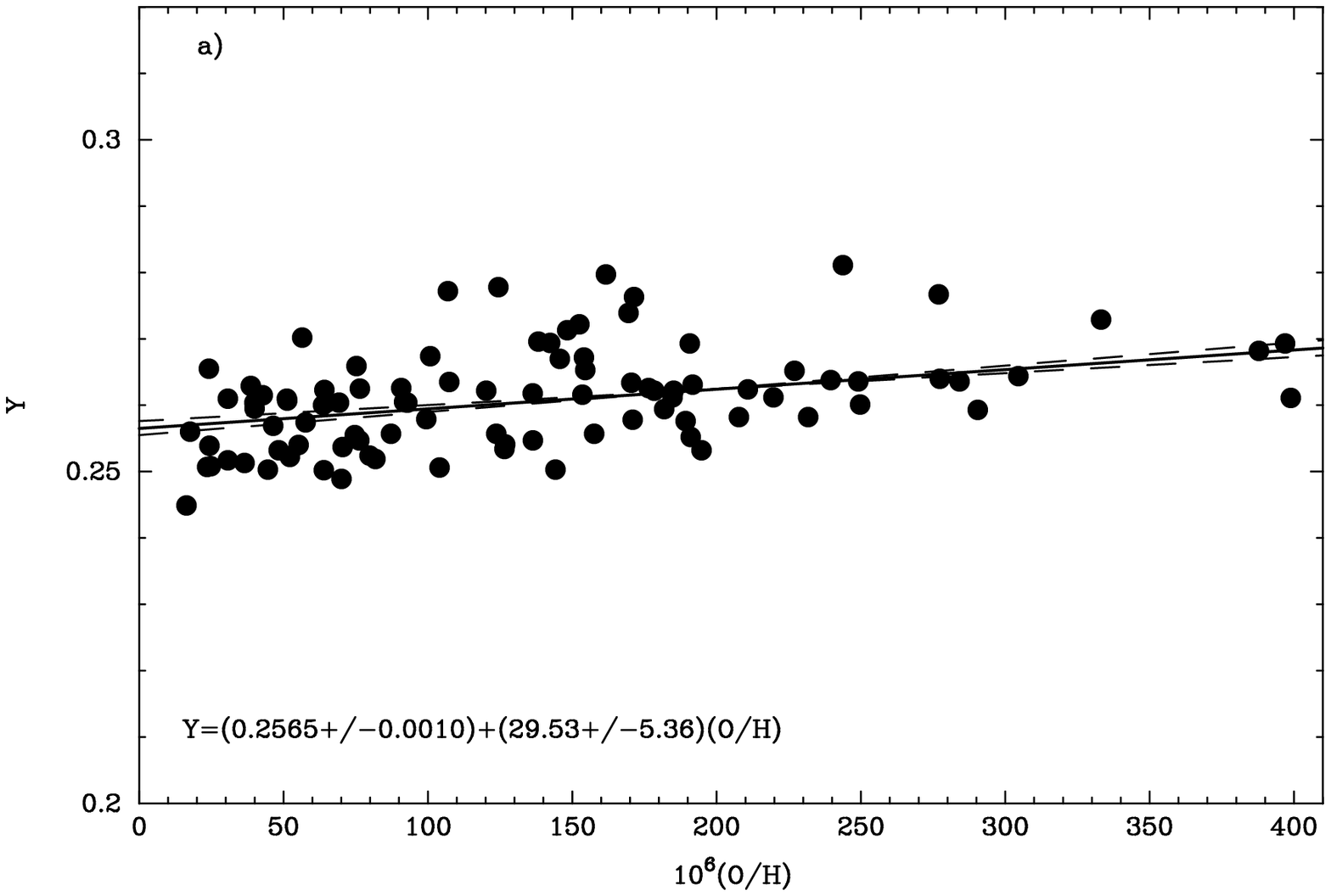}{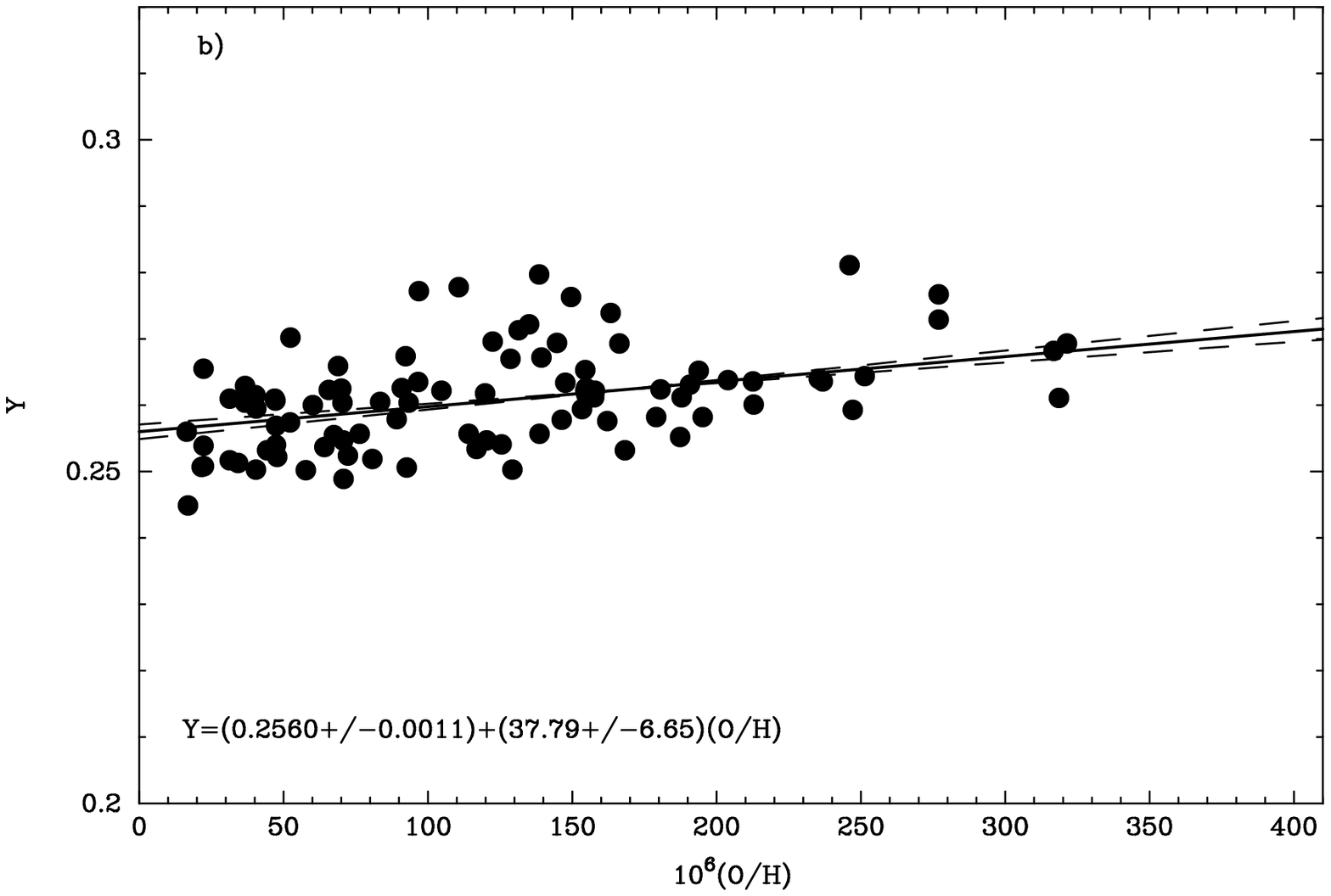}
\figcaption{  Linear regressions of the helium mass fraction $Y$ vs. oxygen 
abundance for H {\sc ii} regions in the HeBCD sample. 
The $Y$s are derived with the He {\sc i} emissivities
from \citet{P05}. The electron temperature 
$T_e$(He$^+$) is varied in the range (0.95 -- 1)$\times$$T_e$(O {\sc iii}).
The oxygen abundance is derived adopting an electron temperature 
equal to $T_e$(He$^+$) in a) and to  
$T_e$(O {\sc iii}) in b).
\label{fig1}}
\end{figure*}

Two $Y$ -- O/H linear regressions
for the HeBCD galaxy sample of \citet{I07}, 
with the above set of parameters, are shown in Fig. \ref{fig1}. 
The two regression lines differ in the way 
oxygen abundances have been calculated. For the 
first regression line (Fig. \ref{fig1}a), 
oxygen abundances have been derived by setting the temperature
of the O$^{++}$ zone equal to $T_e$(He$^+$), 
while for the second 
(Fig. \ref{fig1}b), they have been derived by adopting the temperature  
$T_e$(O {\sc iii}) derived from the 
[O {\sc iii}] $\lambda$4363/($\lambda$4959+$\lambda$5007) line flux ratio.

The primordial values obtained
from the two regressions in Fig. \ref{fig1}, $Y_p$ = 0.2565 $\pm$ 0.0010
and $Y_p$ = 0.2560 $\pm$ 0.0011, are very similar but are 
significantly higher than the
value $Y_p$ = 0.2516 $\pm$ 0.0011 obtained by \citet{I07}
for the same galaxy sample. The 2\% difference is due to the inclusion
of the correction for fluorescent excitation of H lines, the
correction for a larger correction for collisional excitation to the 
H$\beta$ flux and larger adopted equivalent widths of the stellar He {\sc i}
5876, 6678 and 7065 absorption lines.
We adopt the value of $Y_p$ from 
Fig. \ref{fig1}a, 
where both O/H and $Y$ are calculated with the same temperature 
$T_e$ = $T_e$(He$^+$).

We have varied the ranges of some parameters
to study how the value of $Y_p$ is affected by these variations. 
We have found that varying the fraction of  
fluorescent excitation of the hydrogen lines between 0\% and 2\%, and/or 
setting $T_e$(He$^+$) = $T_e$(O {\sc iii}) 
or changing $T_e$(He$^+$) in the range 
(0.9 -- 1.0)$\times$ $T_e$(O {\sc iii}) (instead of making 
it change between  0.95 and 1.0 $\times$ $T_e$(O {\sc iii})),
result in a change of $Y_p$ between 0.254 and 0.258. 
Additionally, adding a systematic error of 1\% caused by 
uncertainties in the He {\sc i} emissivities \citep{P09}, gives 
$Y_p$= 0.2565 $\pm$ 0.0010(stat.) $\pm$ 0.0050(syst.), 
where ``stat'' and ``syst'' 
refer to statistical and systematic errors, respectively.  
Thus, the value of $Y_p$ derived in this paper,
is 3.3\% greater than the 
value of 0.2482 
obtained from the 3yr WMAP data, assuming SBBN \citep{S07}. 
However, it is consistent
with the $Y_p$ = 0.25$^{+0.10}_{-0.07}$ obtained by \citet{Ich08} 
from the available WMAP, ACBAR, CBI, and BOOMERANG data [actually, the
peak value in their one-dimensional marginalized distribution of $Y_p$
(their Fig.3) is equal to 0.254].

Using Eq. \ref{eq:dO}, we derive from the $Y$ -- O/H linear regression
(Fig. \ref{fig1}a) the slopes 
d$Y$/dO = 2.46$\pm$0.45(stat.) and d$Y$/d$Z$ = 1.62$\pm$0.29(stat.).
These slopes are shallower than the ones of 4.33$\pm$0.75 and 
2.85$\pm$0.49 derived by \citet{I07}.

\section{Deviations from SBBN}\label{cosmo}

We now use our derived value of the primordial He abundance along with
the observed primordial abundances of other light elements to check the 
consistency of SBBN.   
Deviations from the standard rate of Hubble expansion in the early Universe 
can be caused by an extra contribution to the total energy density, for 
example by additional flavors of
neutrinos. 
The total number of different species of 
weakly interacting light relativistic particles can be 
conveniently be parameterized by $N_\nu$, 
the ``effective number of light neutrino species''. 

To perform the study, we use the statistical $\chi^2$ technique,
 with the code described by
\citet{Fi98} and \citet{Li99}. This code allows to 
analyze the constraints 
that the measured He, D and $^7$Li abundances put 
on $\eta$ and $N_\nu$. 
For the primordial D abundance, we use the 
value obtained by \citet{P08}. As for $^7$Li, its value derived from
observations of low-metallicity halo stars \citep{As06} is $\sim$ 5
times lower than the one obtained from the WMAP analysis \citep{D09}.
Because mechanisms that may lead to a reduction of the $^7$Li 
primordial abundance, such as diffusion or rotationally induced mixing,
are not well understood and we do not know how to correct for them, we have    
adopted the value of the primordial abundance of $^7$Li abundance 
as derived from the 5yr WMAP data of \citet{D09}.
The predicted primordial abundances of light elements depend
on the adopted neutron life-time $\tau_{\rm n}$. 
We have considered two values,
the old one, $\tau_{\rm n}$ = 885.4 $\pm$ 0.9 s \citep{A00}, and the
new one, $\tau_{\rm n}$ = 878.5 $\pm$ 0.8 s \citep{S05,S08}.
With the old value of $\tau_{\rm n}$ and our best 
value of the primordial He abundance, $Y_p$ = 0.2565$\pm$ 0.0010(stat.) $\pm$ 0.0050(syst.),
the minimum $\chi^2_{min}$ (= 0.640524) is obtained
when $\eta_{10}$ = 6.47 and $N_\nu$ = 3.68. The value of $\eta_{10}$ is in
agreement with $\eta_{10}$ = 6.23$\pm$0.17 derived from the WMAP data 
\citep{D09}. 
If instead the new value of $\tau_{\rm n}$ 
is adopted with the same value of  $Y_p$, then 
the minimum $\chi^2_{min}$ (= 0.619816) is obtained
when $\eta_{10}$ = 6.51 and $N_\nu$ = 3.80. We note that $\eta_{10}$ and
$N_\nu$ only slightly depend on the value of the $^7$Li abundance. They are
decreased by 3\% and 2\%, respectively, if the
observed $^7$Li abundance by \citet{As06} is adopted.
 
\begin{figure*}
\figurenum{2}
\epsscale{1.1}
\plottwo{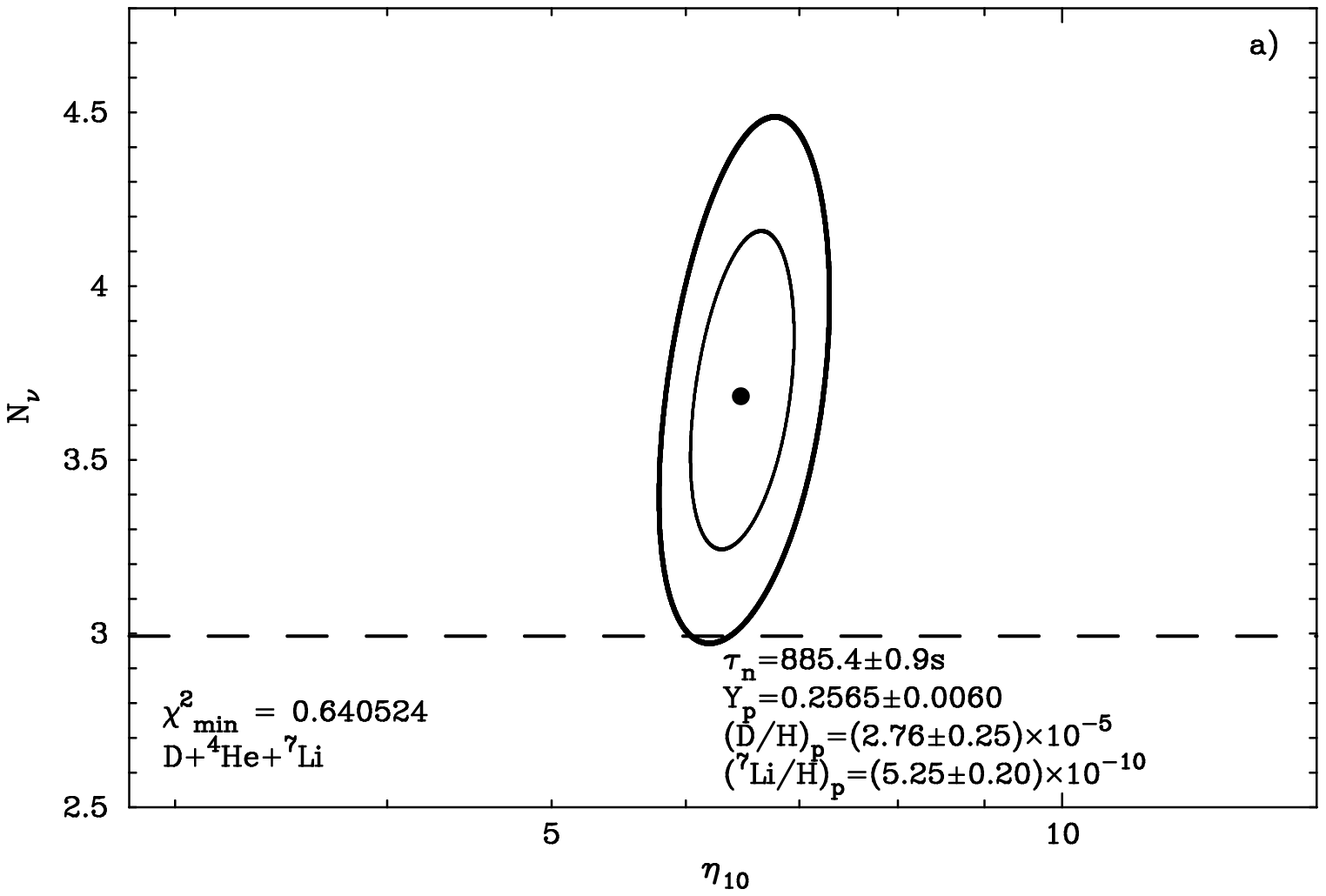}{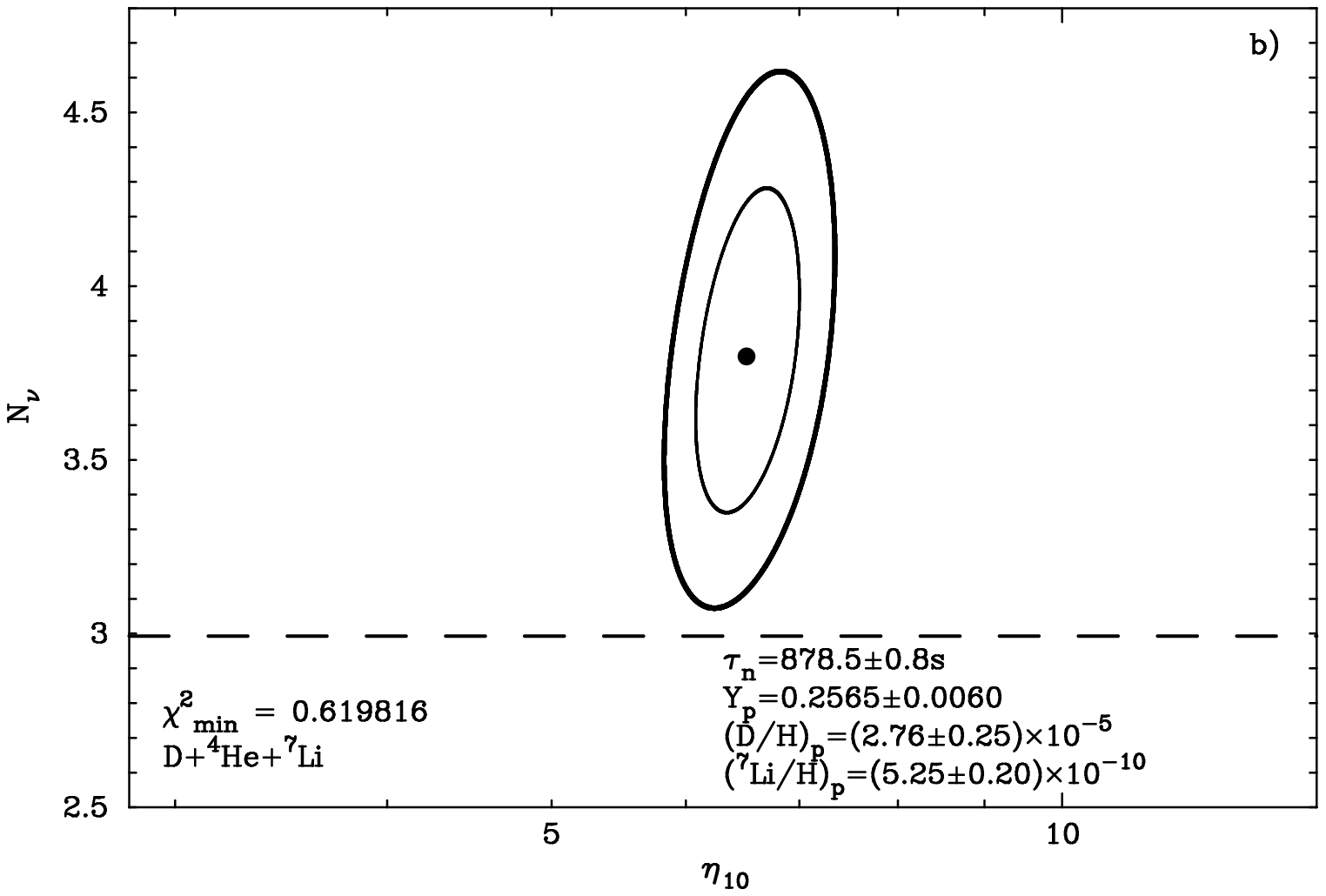}
\figcaption{a) Joint fits to the baryon-to-photon number ratio, 
$\eta_{10}$, and the equivalent number of light neutrino species $N_\nu$, 
using a $\chi^2$ analysis with the code developed by \citet{Fi98}.
and \citet{Li99}. The value of the primordial He abundance has been set to 
$Y_p$ = 0.2565 (this paper), 
that of (D/H)$_p$ is taken from \citet{P08} and that of 
($^7$Li/H)$_p$ from 5yr WMAP measurements \citet{D09}.
A neutron lifetime $\tau_{\rm n}$ = 885.4 $\pm$ 0.9s from \citet{A00}
has been adopted.
Thin and thick solid lines represent 
respectively 1$\sigma$ and 2$\sigma$ deviations.
The experimental value $N_\nu$ = 2.993 \citep{Ca98} is shown by the
dashed line. b) The same as in (a), but with a neutron lifetime 
$\tau_{\rm n}$ = 878.5 $\pm$ 0.8s \citep{S05,S08}.
\label{fig14}
\label{fig2}}
\end{figure*}

The joint fit of $\eta$ and $N_\nu$ is shown in Figures \ref{fig2}a
and \ref{fig2}b for the two values of $\tau_{\rm n}$.
The 1$\sigma$ ($\chi^2$ -- $\chi^2_{min}$ = 1.0) and
2$\sigma$ ($\chi^2$ -- $\chi^2_{min}$ = 2.71) deviations are shown 
respectively by the thin and thick solid lines. 
We find the equivalent number of light neutrino species to be in the range
$N_\nu$ = 3.68$^{+0.80}_{-0.70}$ (2$\sigma$) (Fig. \ref{fig2}a) in the 
first case, and
$N_\nu$ = 3.80$^{+0.80}_{-0.70}$ (2$\sigma$) (Fig. \ref{fig2}b) in the second 
case. 
Both of these values are only marginally consistent (at the 2$\sigma$ level) 
with the experimental value of 2.993$\pm$0.011 \citep{Ca98} shown by the 
dashed line, implying deviations from 
SBBN.  
We note that, although both values are consistent with 
$N_\nu$ = 4.4 $\pm$ 1.5 derived from the analysis
of 5yr WMAP observations \citep{K09}, 
the primordial helium abundance sets tighter
constraints on the effective number of neutrino species than  
the CMB data, the error bars of $N_\nu$
being approximately half as large in the first case as compared to  
the latter case.

\section{SUMMARY AND CONCLUSIONS}\label{summary}

     We present here a new determination of the primordial
helium mass fraction $Y_p$ by linear regressions of a sample of 93 spectra
of 86 low-metallicity extragalactic H {\sc ii} regions. 

In this new determination of $Y_p$, we have taken into account the 
latest developments concerning several   
known systematic effects. 
We have used Monte Carlo methods to solve simultaneously for 
the effects of collisional and fluorescent enhancements 
of He {\sc i} recombination lines, of
collisional and fluorescent excitation of hydrogen emission lines, 
of underlying stellar
He {\sc i} absorption, of possible temperature 
differences between the He$^+$ and [O {\sc iii}] zones, 
and of the ionization correction factor $ICF$(He$^+$ + He$^{2+}$). 

We have obtained the following results:

1. Our best value is $Y_p$ = 0.2565 $\pm$ 0.0010(stat.) $\pm$ 0.0050(syst.), 
or 3.3\% larger than the value derived from the 
microwave background radiation fluctuation measurements assuming SBBN. 
In order to bring this high value of 
$Y_p$ into agreement with the deuterium and 5yr WMAP measurements, 
an equivalent number of neutrino flavors in the range 3.68 -- 3.80,
depending on the lifetime of the neutron, is required. 
This is higher than the canonical value of 3 and  
implies the existence of deviations from SBBN. 

2. The $dY/dZ$ slope derived from the $Y$ -- O/H linear
regression is equal to 1.62 $\pm$ 0.29(stat.), shallower than the previous
determination by \citet{I07}.

\acknowledgements
Y.I.I. thanks the staff of the Astronomy Department at the
University of Virginia and of the Max Planck Institute for Radioastronomy 
in Bonn, Germany for warm hospitality. 



\begin{thebibliography}{}

\bibitem[Aller(1984)]{A84} Aller, L. H. 1984, Physics of Thermal
Gaseous Nebulae (Dordrecht: Reidel)
\bibitem[Arzumanov et al.(2000)]{A00} Arzumanov, S., Bondarenko, L., 
Chernyavsky, S., et al. 2000, Physics Letters B 483, 15
\bibitem[Asplund et al.(2006)]{As06} Asplund, M., Lambert, D. L., 
Nissen, P. E., Primas, F., \& Smith, V. V. 2006, \apj, 644, 229
\bibitem[Benjamin et al.(1999)]{B99} Benjamin, R. A.,
Skillman, E. D., \& Smits, D. P. 1999, \apj, 514, 307
\bibitem[Benjamin et al.(2002)]{B02} Benjamin, R. A.,
Skillman, E. D., \& Smits, D. P. 2002,
\apj, 569, 288
\bibitem[Cardelli et al.(1989)]{C89} Cardelli, J. A., Clayton, G. C., \&
Mathis, J. S. 1989, \apj, 345, 245
\bibitem[Caso et al.(1998)]{Ca98} Caso, C., et al. 
(Particle Data Group) 1998, Eur. J. Phys., C3, 1
\bibitem[Dunkley et al.(2009)]{D09} Dunkley, J., et al. 2009, \apjs, 180, 306
\bibitem[Fiorentini et al.(1998)]{Fi98} Fiorentini, G., Lisi, E., Sarkar, S., 
\& Villante, F. L. 1998, \prd, 58, 063506
\bibitem[Fukugita \& Kawasaki(2006)]{F06} Fukugita, M., \& Kawasaki, M.
2006, \apj, 646, 691
\bibitem[Gonz\'alez Delgado et al.(2005)]{G05} Gonz\'alez Delgado, R. M., 
Cervi\~no, M., Martins, L. P., 
Leitherer, C., \& Hauschildt, P. H. 2005, \mnras, 357, 945
\bibitem[Guseva et al.(2006)]{G06} Guseva, N. G., 
Izotov, Y. I., \& Thuan, T. X. 2006, \apj, 644, 890
\bibitem[Guseva et al.(2007)]{G07} Guseva, N. G., Izotov, Y. I., 
Papaderos, P., \& Fricke, K. J. 2007, \aap, 464, 885
\bibitem[Ichikawa et al.(2008)]{Ich08} Ichikawa, K., Sekiguchi, T., 
\& Takahashi, T. 2008, \prd, 78, 043509
\bibitem[Izotov \& Thuan(2004)]{IT04} Izotov, Y. I., \& Thuan, T. X. 2004, \apj, 602, 200
\bibitem[Izotov et al.(2006)]{I06} Izotov, Y. I., Stasi\'nska, G., Meynet, G.,
Guseva, N. G., \& Thuan, T. X. 2006, \aap, 448, 955
\bibitem[Izotov et al.(2007)]{I07} Izotov, Y. I., Thuan, T. X., \&
Stasi\'nska, G. 2007, \apj, 662, 15
\bibitem[Komatsu et al.(2009)]{K09} Komatsu, E., et al. 2009, \apjs, 180, 330
\bibitem[Leitherer et al.(1999)]{L99} Leitherer, C., Schaerer, D.,
Goldader, J. D., et al. 1999, \apjs, 123, 3
\bibitem[Lisi et al.(1999)]{Li99} Lisi, E., Sarkar, S., \& Villante, F. L. 
1999, \prd, 59, 123520
\bibitem[Luridiana(2009)]{L09a} Luridiana, V. 2009, \apss, 324, 361
\bibitem[Luridiana et al.(2009)]{L09b} Luridiana, V., Sim\'on-D\'iaz, S.,
Cervi\~no, M., et al. 2009, \apj, 691, 1712
\bibitem[Pagel et al.(1992)]{P92} Pagel, B. E. J., Simonson, E. A.,
Terlevich, R. J., \& Edmunds, M. G. 1992,
\mnras, 255, 325
\bibitem[Peimbert \& Torres-Peimbert(1974)]{PTP74} Peimbert, M., \&
Torres-Peimbert, S. 1974, \apj, 193, 327
\bibitem[Peimbert \& Torres-Peimbert(1976)]{PTP76} Peimbert, M., \&
Torres-Peimbert, S. 1976, \apj, 203, 581
\bibitem[Peimbert et al.(2007)]{P07} Peimbert, M.,
Luridiana, V., \& Peimbert, A. 2007, \apj, 666, 636
\bibitem[Pettini et al.(2008)]{P08} Pettini, M., Zych, B. J., Murphy, M. T., 
Lewis, A., Steidel, C. C. 2008, \mnras, 391, 1499
\bibitem[Porter et al.(2005)]{P05} Porter, R. L., Bauman, R. P., 
Ferland, G. J., \& MacAdam, K. B. 2005, \apj, 622, L73
\bibitem[Porter et al.(2009)]{P09} Porter, R. L., Ferland, G. J., 
MacAdam, K. B., \& Storey, P. J. 2009, \mnras, 393, L36
\bibitem[Press et al.(1992)]{Pr92} Press, W. H., Teukolsky, S. A., 
Vetterling, W. T.,\& Flannery, B. P., 1992, Numerical Recipes in C, The Art of 
Scientific Computing /Second Edition/, Cambrige University Press
\bibitem[Serebrov et al.(2005)]{S05} Serebrov, A. P., Varlamov, V. E., 
Kharitonov, A. G., et al. 2005, Physics Letters B 605, 72
\bibitem[Serebrov et al.(2008)]{S08} Serebrov, A. P., Varlamov, V. E., 
Kharitonov, A. G., et al. 2008, Phys. Rev. C 78, 035505
\bibitem[Spergel et al.(2007)]{S07} Spergel, D. N., Bean, R., Dor\'e, O.,
et al. 2007, \apjs, 170, 377
\bibitem[Stasi\'nska \& Izotov(2001)]{SI01} Stasi\'nska, G., \& Izotov, Y. I.
2001, \aap, 378, 817
\bibitem[Steigman(2006)]{St06}  Steigman, G. 2006, Int. J. Mod. Phys. E, 
15, 1
\bibitem[Steigman(2007)]{St07}  Steigman, G. 2007, Annual Review of Nuclear 
and Particle Science, 57, 463
\bibitem[Whitford(1958)]{W58} Whitford, A. E. 1958, \aj, 63, 201
\end{thebibliography}
\end{document}